# MATCHING SCHERRER'S K ESSENCE ARGUMENT WITH ALTERATIONS OF DI QUARK SCALAR FIELDS PERMITTING AN EVENTUAL COSMOLOGICAL CONSTANT DOMINATED INFLATIONARY EXPANSION

A. W. Beckwith

## ABSTRACT

We previously showed that we can use di-quark pairs as a model of how nucleation of a new universe occurs. We now can construct a model showing evolution from a dark matter dark energy mix to a pure cosmological constant cosmology due to changes in the slope of the resulting scalar field, using much of Scherrer's k-essence model. This same construction permits a use of the speed of sound, in k essence models evolving from zero to one. Having the sound speed eventually reach unity permits matching conventional cosmological constant observations in the aftermath of change of slope of a di-quark pair generated scalar field during the nucleation process of a new universe. These results are consistent with applying Bunyi and Hus semi classical criteria for cosmological potentials to indicate a phase transition alluded to by Dr. Edward Kolbs model of how the initial degrees of freedom declined from over 100 to something approaching what we see today in flat space cosmology

Correspondence: A. W. Beckwith:     projectbeckwith2@yahoo.com

PACS numbers: 03.75.Lm, 11.27.+d, 98.65.Dx, 98.80.Cq, 98.80.-k

# INTRODUCTION.

As of June 2005, an effort was made to combine reconstruction of data gathering techniques with the requirement of the JDEM dark matter-dark energy search for the origins of dark matter in the early universe[1]. We are now determining the role initial dark matter plays in the evolution of di quark states to conventional chaotic inflation We have initial di quark scalar fields in a near perfect thin wall approximation moving toward the Gaussian wave functional used in the ground state of the harmonic oscillator, with obvious connections to chaotic inflation used by Guth[2], and others. In particular, our wave functionals used for the di quark scalar field[3]s are consistent with evolution from a perturbed wash board potential system potential, to the chaotic inflationary potential used by Guth[2]

We[4] have investigated the role an initial false vacuum procedure with a perturbed Washboard (Sine Gordon) potential plays in the nucleation of a scalar field in inflationary cosmology. Here, we show how that same scalar field permits cosmological state evolution from a thin wall di quark scalar field state of matter into the chaotic inflationary cosmology presented by Guth[2]. The final results of this model coincides with a region that matches the flat slow roll requirement of $\left|\frac{\partial^2 V}{\partial \phi^2}\right| << H^2$ that is a requirement of realistic inflation models. Doing so leads to initial matter states- analyzed via k essence methodology, by using scalar fields which are congruent to known potential systems.

We have formed a model, using Scherrer's argument, [5] evaluating initial conditions to shed light on if this model universe is radiation-dominated in the beginning or congruent with assuming an Einstein cosmological constant driving force for expansion. We do not have a cosmological constant-dominated era when our scalar field φ is close to a thin wall approximation . The sharpness of this slope, leading to a near delta function for derivatives of the scalar field φ for the thin wall approximation for the initial conditions of an expanding universe would void a role for an Einstein cosmological constant unless there was a later deterioration of the thin wall approximation.

## HOW DARK MATTER TIES IN, USING PURE KINETIC K ESSENCE AS DARK MATTER TEMPLATE FOR A NEAR THIN WALL APPROXIMATION OF φ

We define k essence as any scalar field with non-cannonical kinetic terms. Following Scherrer,[5] we introduce a momentum expression via

$$p = V(\phi) \cdot F(X) \tag{1}$$

where we set[3] $X$ as a scalar 'derivative' and present $F$ as

$$X = \frac{1}{2} \cdot \nabla_\mu \phi \ \nabla^\mu \phi \tag{2}$$

$$F = F_0 + F_2 \cdot (X - X_0)^2 \equiv F_0 + F_2 \cdot \varepsilon \tag{3}$$

where we define $X_0$ via $F_X\big|_{X=X_0} = \dfrac{dF}{dX}\bigg|_{X=X_0} = 0$, as well as use

$$w \equiv \frac{p}{\rho} \equiv \frac{F}{2 \cdot X \cdot F_X - F} \tag{4}$$

as well as a growth of density perturbations terms factor Garriga and Mukhanov[6] wrote as

$$C_S^2 = \frac{(\partial p/\partial X)}{(\partial \rho/\partial X)}\bigg|_{X=s} \equiv \frac{F_X}{F_X + 2 \cdot X \cdot F_{XX}} \cong \frac{1}{1 + 2 \cdot (X_0 + \tilde{\varepsilon}_0) \cdot (1/\tilde{\varepsilon}_0)} \tag{5}$$

where,

$$(F_X + 2 \cdot X \cdot F_{XX}) \cdot \ddot{X} + 3 \cdot H \cdot F_X \cdot \dot{X} \cong 0 \qquad (6)$$

and when

$$(\partial_\mu \phi) \cdot (\partial^\mu \phi) \equiv \left(\frac{1}{c} \cdot \frac{\partial \phi}{\partial \cdot t}\right)^2 - (\nabla \phi)^2 \cong -(\nabla \phi)^2 \to -\left(\frac{d}{dx} \phi\right)^2 \qquad (7)$$

We get these values for phase being a box of height scaled to be $2 \cdot \pi$ and of width $L$.

$$\phi \approx \pi \cdot [\tanh b \cdot (x + L/2) - \tanh b \cdot (x - L/2)] \qquad (8)$$

and looking at what happens when we look at either

## CASE I

$$|X_0| \approx \frac{1}{2} \cdot \left(\frac{\partial \phi}{\partial x}\right)^2 \cong \frac{1}{2}[\delta_n^2(x + L/2) + \delta_n^2(x - L/2)] \gg \tilde{\varepsilon}_0, \neq \infty \qquad (9)$$

when $n \longrightarrow$ large value N ,

or

## CASE II

$n \to \infty$ leading directly to

$$|X_0| \approx \frac{1}{2} \cdot \left(\frac{\partial \phi}{\partial x}\right)^2 \cong \infty \qquad (11)$$

Or

## CASE III

$$|X_0| \approx \frac{1}{2} \cdot \left(\frac{\partial \phi}{\partial x}\right)^2 \leq \tilde{\varepsilon}_0 \qquad (12)$$

These will be tabulated in a table below, and commented upon physically speaking.

# SUMMARIZING DIFFERENT SITUATIONS WITH RESPECT TO DI QUARK PAIR THIN WALL CONTRIBUTIONS TO K ESSENCE VALUES

| $X_0$ | $C_S^2$ | $w$ |
|---|---|---|
| $\gg \varepsilon_0 \geq 0$ | $0 \leq \varepsilon^+ \ll 1$ | 0,-1 |
| $\infty$ | 0 | 0 |
| $\varepsilon_0 \geq 0$ | 1 | -1 |

## CASE I

is for when $b \geq 10$ in Eq (8) we can obtain ( depending upon what we pick for initial conditions)

and recover Sherrer's solution for the speed of sound[5]

$$C_S^2 \approx \frac{1}{1 + 4 \cdot X_0 \left(1 + \frac{X_0}{2 \cdot \tilde{\varepsilon}_0}\right)} \to 0 \tag{13}$$

But we also find that there is a continuum of different values

$$w \equiv \frac{p}{\rho} \cong \frac{-1}{1 - 4 \cdot (X_0 + \tilde{\varepsilon}_0) \cdot \left(\frac{F_2}{F_0 + F_2 \cdot (\tilde{\varepsilon}_0)^2} \cdot \tilde{\varepsilon}_0\right)} \approx \frac{-1}{1 - 4 \cdot \frac{X_0 \cdot \tilde{\varepsilon}_0}{F_2}} \to 0, -1 \tag{14}$$

This means that the initial conditions if an example $F_2 \to 10^3$, $\tilde{\varepsilon}_0 \to 10^{-2}$, $X_0 \to 10^3$,

$\Rightarrow w \equiv \frac{p}{\rho} \approx -1$ which is necessary for cosmological inflation We have to settle for dark matter dominated cosmological expansion due to $C_S^2 \approx 0$ zero effective sound speed. When $w \approx 0$ we do not have inflationary expansion, The 1st case is consistent with a near thin wall di quark pairs with $w \approx -1$ denoting cosmological expansion, and $w \approx 0$ discarded as unphysical.

## CASE II

is for when when $b >> 10$ in Eq (8) $\Leftrightarrow |X \approx X_0| \to \infty$ if the ensemble of S-S' di quark pairs were represented by a pure thin wall approximation $\delta_n(x \pm L/2) \xrightarrow[n \to \infty]{} \delta(x \pm L/2)$. This is unphysical and is discarded. Due to $C_S^2 \approx 0$, and $w \cong 0$ at the same time.

## CASE III

is for when when $b \geq 3$ in Eq (8) we obtain

$$w \cong \frac{-1}{1 - 4 \cdot \frac{X_0 \cdot \tilde{\varepsilon}_0}{F_2}} \to -1 \qquad (15)$$

and

$$C_S^2 \approx \frac{1}{1 + 4 \cdot X_0 \left(1 + \frac{X_0}{2 \cdot \tilde{\varepsilon}_0}\right)} \to 1 \qquad (16)$$

(17)

This eliminates a pure radiation state initial condition (as Cardone et al postulated), i.e. we have universe as being first radiation dominated, then dark matter dominated, and finally dark energy. The wave functional associated with this as a non thin wall di quark scalar field. This di quark scalar model is substantially similar to the soliton-anti soliton pairs used in nucleated states used in condensed matter transport involving density wave transport[7]. It also is a sutiable candidate for dark matter-dark energy as noted by Zhitinisky[8] Finally, this also reflects an Einstein cosmological constant dominated inflationary expansion.

## CONCLUSION

We have a situation for which we can postulate an early universe which is *not* necessarily radiation dominated as postulated by Carbone et al. We should keep in mind that Sherrer was looking for very small $\varepsilon_1$ and a constant $a_1 > a$, with $a$ written as an expansion scale factor[5].

$$X = X_0 \cdot \left(1 + \varepsilon_1 \cdot \left(\frac{a}{a_1}\right)^{-3}\right) \tag{18}$$

so he could then get a *general* solution of

$$1 \gg C_x^2 \equiv (X - X_0)/(3 \cdot X - X_0) \equiv \frac{1}{2} \cdot \varepsilon_1 \cdot (a/a_1)^{-3} \approx \varepsilon^+ \geq 0 \tag{19}$$

while at the same time keeping $w = -1$.

Our Case I above fits this scenario. Our Case III has $C_x^2 \cong 1, w \equiv -1$, which is what we need for an Einstein constant dominated cosmological expansion. Our Case II iunphysical, with both $C_x^2, w \to 0$, and we need density and pressure as negative reciprocals of each other to have a realistic cosmological inflationary expansion. What we are doing is indicating a bridge between an Einstein constant cosmology, and initial states of a dark matter-dark energy dominated era

In addition, our kinetic model can be compared with the very interesting Chimentos[9] purely kinetic k–essence model, with density fluctuation behavior The model we are using indicates our density functional reach $\rho =$ constant after passing through the tunneling barrier implicit in the formation of di quark pairs, as analogous to S-S' pairs used in condensed matter models[7] This is similar to the density fluctuations and uncertainty principle ideas used as early as 1992, for analyzing early universe

states[10].Furthermore, the methodology presented makes full use of the idea embodied in a condensed matter application of tunneling Hamiltonians for current calculations done in my dissertation, and published in IJMPB, where wave functional states formed to match initial conditions to the false vacuum generating perturbed wash board potential are used for kinetic dynamics of charge transport. I.e. the thin wall approximation of a S-S' pair are embodied in the di-quark model initially pioneered by Zhitinisky in 2002, using the so called QCD balls for soliton( anti soliton) states are set up and inserted directly into the Scherrer k essence formalism, there by by passing the necessity of putting in a derivative of the potential term in Eq (6) above. All the above permits including in the effects of a first order phase transformation from a false vacuum potential which is used to get thin wall approximations, to the Gaussian potential formed situation with no thin wall approximation.

Another calculation, shown in **Appendix I** involving a semi classical approximation breaking down after going from a washboard potential slightly perturbed by an an harmonic term, to the Gaussian wave functional congruent potential for chaotic inflation written by Guth also indicates the existence of a phase transition The semi classical approximation holding for thin wall di quark models is for the perturbed Washboard potential, while the semi classical approximation breaking down for the Guth style chaotic This is what is observed in **figures 1a to 3** below

All of this argues that the initial situation transforming from a dark matter-dark energy initial state of cosmology, to one with a cosmology dominated by the Einstein cosmological constant involves a phase transition, which is also in sync with Dr. Edward Kolbs observation about a drop in the initial degrees of freedom of an early cosmology[11,12]. This also is relevant to the assumed initial six dimensions of space in dark matter dominated spatial force interactions, which shows up as a factor in **Appendix I**s treatment of the semi classical approximation. As shown in the table in the bottom of that appendix, six dimensions, and dark matter dominated conditions[13] are de rigor for early cosmology, while the Euclidian flat space approximation used at the end of chaotic inflation as described by Guth involve far fewer dimensions.

# FIGURE CAPTIONS

**Fig 1a, 1b**: Evolution of the phase from a thin wall approximation to a more nuanced thicker wall approximation with increasing L between di quark S-S' instanton componets. The height drops and the width L increases corresponds to a de evolution of the thin wall approximation. This is in tandem with a collapse to Guth's[2] standard chaotic scalar $\phi^2$ potential system. As the thin wall approximation dissipates, the physical system approaches standard cosmological constant behavior.

**Fig. 2a, 1b**: As the *walls* of the S-S' pair approach the thin wall approximation, one finds that for a normalized distance $L = 9 \to L = 6 \to L = 3$ that one has an approach toward delta function behavior at the boundaries of the new, nucleating phase. As *L* increases, the delta function behavior subsides dramatically.

**Fig 3**: Initial configuration of the domain wall nucleation potential as given by appendix I Eq. (4a) which we claim eventually becomes in sync with appendix I Eq. (4c) due to the phase transition alluded to by Dr. Edward Kolbs model of how the initial degrees of freedom declined from over 100 to something approaching what we see today in flat Euclidian space models of space time (i.e. the FRW metric used in standard cosmology)

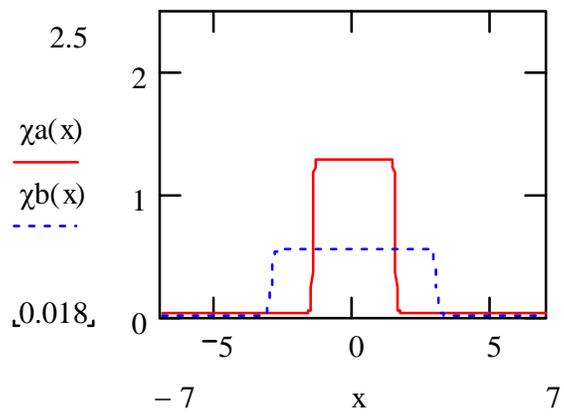

Figure 1a

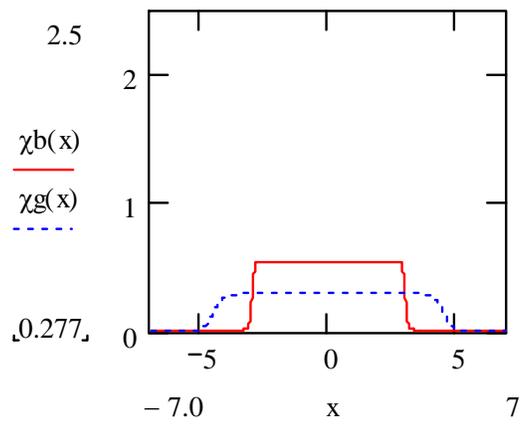

Figure 1b

Beckwith

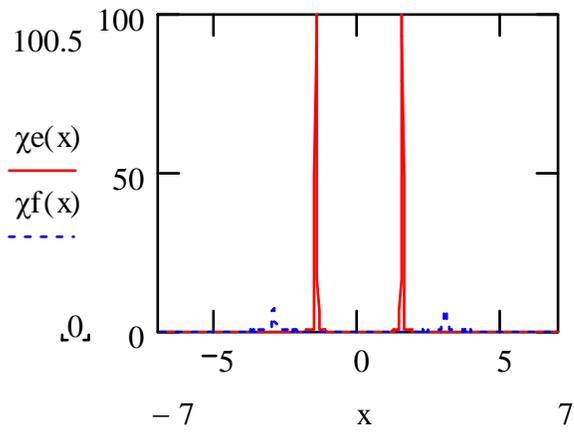

Figure 2a

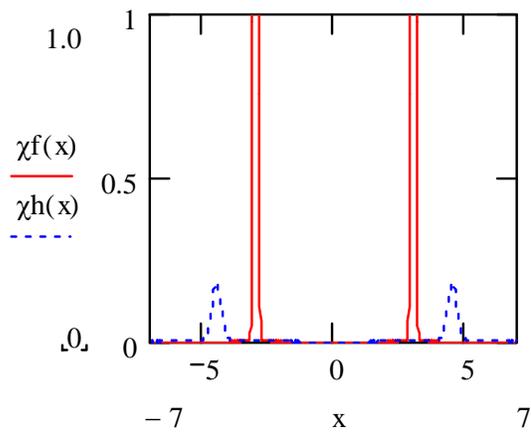

Figure 2b

Beckwith

Figure 3

Beckwith

# APPRENDIX I. HOW TO ANALYZE PHYSICAL STATES IN THE PRECURSORS TO INFLATIONARY COSMOLOGY

Let us first consider an elementary definition of what constitutes a semi classical state. As visualized by Buniy and Hsu,[14] it is of the form $|a\rangle$ which has the following properties:

i) Assume $\langle a|1|a\rangle = 1$

(Where 1 is an assumed identity operator, such that $1|a\rangle = |a\rangle$)

ii) We assume that $|a\rangle$ is a state whose probability distribution is peaked about a central value, in a particular basis, defined by an operator $Z$

a) Our assumption above will naturally lead, for some $n$ values

$$\langle a|Z^n|a\rangle \equiv (\langle a|Z|a\rangle)^n \tag{1}$$

Furthermore, this will lead to, if an operator $Z$ obeys Eq. (1) that if there exists another operator, call it $Y$ which does not obey Eq. (1), that usually we have non commutativity

$$[Y,Z] \neq 0 \tag{2}$$

Buniy and Hsu[14] speculate that we can, in certain cases, approximate a semi classical evolution equation of state for physical evolution of cosmological states with respect to classical physics operators. This well may be possible for post inflationary cosmology; however, in the initial phases of quantum nucleation of a universe, it does not apply.

To review our model of S-S' pair nucleosynthesis for di quark pair states in an early universe, first is the issue of how the potential evolved. Namely:

$$V_1 \to V_2 \to V_3$$
$$\phi(increase) \leq 2\cdot\pi \to \phi(decrease) \leq 2\cdot\pi \to \phi \approx \varepsilon^+ \qquad (3)$$
$$t \leq t_P \qquad \to t \geq t_P + \delta\cdot t \qquad \to t >> t_P$$

We described the potentials $V_1$, $V_2$, and $V_3$ in terms of S-S' di quark pairs nucleating and then contributing to a chaotic inflationary scalar potential system.

$$V_1(\phi) = \frac{M_P^2}{2}\cdot(1-\cos(\phi)) + \frac{m^2}{2}\cdot(\phi-\phi^*)^2 \qquad (4a)$$

$$V_2(\phi) \approx \frac{(1/2)\cdot m^2\phi^2}{(1+A\cdot\phi^3)} \qquad (4b)$$

$$V_3(\phi) \approx (1/2)\cdot m^2\phi^2 \qquad (4c)$$

Note that Eq. (4a) has a measure of the onset of quantum fluctuations[2]

$$\phi^* \equiv \left(\frac{3}{16\cdot\pi}\right)^{\frac{1}{4}} \cdot \frac{M_P^{3/2}}{m^{\frac{1}{2}}}\cdot M_P \to \left(\frac{3}{16\cdot\pi}\right)^{\frac{1}{4}} \cdot \frac{1}{m^{\frac{1}{2}}} \qquad (4d)$$

and should be seen in the context of the fluctuations having an upper bound specified by[2]

$$\tilde{\phi}_0 > \sqrt{\frac{60}{2\cdot\pi}} M_P \approx 3.1 M_P \qquad (4e)$$

Also, the fluctuations Guth[2] had in mind were modeled via

$$\phi \equiv \tilde{\phi}_0 - \frac{m}{\sqrt{12\cdot\pi\cdot G}}\cdot t \qquad (4f)$$

This is for his chaotic inflation model using his potential; which we call the third potential in Eq. (4c)

However, I show elswhere[15] that for the false vacuum hypothesis to hold for Eq. (4a) that there is

$$V_1(\phi_F) - V_1(\phi_T) \cong .373 \propto L^{-1} \cong \alpha \tag{4g}$$

Let us now view a toy problem involving use of a S-S' and [2]

$$\phi \equiv \pi \cdot [\tanh b(x - x_a) + \tanh b(x_b - x)] \tag{5}$$

We can, in this give an approximate wave function as given by:

$$\psi \cong c_1 \cdot \exp(-\tilde{\alpha} \cdot \phi(x)) \tag{6}$$

Then we can look to see if for either D=0(3 dimensional space) or D=3(6 dimensional space) we have[11]

$$\left( \int_{x_a}^{x_b} \psi \cdot V_i \cdot \psi \cdot 4\pi \cdot x^{2+D} \cdot dx \right)^N = \int_{x_a}^{x_b} \psi \cdot [V_i]^N \cdot \psi \cdot 4 \cdot \pi \cdot x^{2+D} \cdot dx \Bigg|_{i=1,2,3} \tag{7}$$

For the first potential system, if we set xb=1, xa= - 1, and b = 10. (a sharp slope) for the scalar field boundary we have.

$$\alpha := \frac{.373}{1} \tag{8}$$

This assumes a Gaussian wave functional of Eq(6)

As well as a power parameter of $v = 9$

The first potential system is re scaled as

$$V1(x) := \frac{1}{2} \cdot (1 - \cos(\phi(x))) - \frac{1}{200} \cdot (\phi(x) - \pi)^2 \tag{9}$$

In addition, the following is used as a rescaling of the inner product

$$c1 := \frac{1}{\int_{-30}^{30} (\exp(-\alpha \cdot \phi(x)))^2 \cdot \frac{\pi^3}{3} \cdot x^5 \, dx} \tag{10}$$

$$c2 := \int_{-30}^{30} (\exp(-\alpha \cdot \phi(x)))^2 \cdot \frac{\pi^3}{3} \cdot x^5 \cdot (V1(x))^\nu \cdot |c1|\, dx \tag{11}$$

$$c3 := \left[ \int_{-30}^{30} (\exp(-\alpha \cdot \phi(x)))^2 \cdot \frac{\pi^3}{3} \cdot x^5 \cdot V1(x) \cdot |c1|\, dx \right]^\nu \tag{12}$$

$$c3b := \frac{c2}{c3} \tag{13}$$

Here,

$$C3b = .999 \tag{13a}$$

For the 2$^{nd}$ potential system, if we assume a sharp slope, i.e. b1 = b = 10, and

$$V2(x) := \frac{1}{2} \cdot \frac{(\phi a(x))^2}{1 + .000001(\phi a(x))^3} \tag{14}$$

If

$$\phi a(x) := \pi \cdot [\, \tanh[b1 \cdot (x - xa)] - \tanh[b1 \cdot (xb - x)]\, ] \tag{15}$$

and a modification of the 'Gaussian width' to be

$$\alpha 1 := \frac{.373}{30} \tag{16}$$

We do specify a denominator, due to a normalization contribution we write as

$$c1a := \frac{1}{\displaystyle\int_{-30}^{30} (\exp(-\alpha 1 \cdot \phi a(x)))^2 \cdot \frac{\pi^3}{3} \cdot x^5\, dx} \tag{17}$$

$$c4 := \int_{-30}^{30} (\exp(-\alpha 1 \cdot \phi a(x)))^2 \cdot \frac{\pi^3}{3} \cdot x^5 \cdot (V2(x))^\nu \cdot |c1a| \, dx \tag{18}$$

In addition:

$$c5 := \left[ \int_{-30}^{30} (\exp(-\alpha \cdot \phi a(x)))^2 \cdot \frac{\pi^3}{3} \cdot x^5 \cdot V2(x) \cdot |c1a| \, dx \right]^\nu \tag{19}$$

We then use a ratio of

$$c5b := \frac{c4}{c5} \tag{20}$$

Here, when one has the six dimensions, plus the thin wall approximation:

$$C5b = 2.926E\text{-}3 \tag{21}$$

When one has three dimensions, plus the thin wall approximation

$$c6 := \int_{-30}^{30} (\exp(-\alpha 1 \cdot \phi a(x)))^2 \cdot \frac{\pi^1}{.25} \cdot x^2 \cdot (V2(x))^\nu \cdot |c1b| \, dx \tag{22}$$

$$c7 := \left[ \int_{-30}^{30} (\exp(-\alpha \cdot \phi(x)))^2 \cdot \frac{\pi^1}{.25} \cdot x^2 \cdot V2(x) \cdot |c1b| \, dx \right]^\nu \tag{23}$$

$$c7b := \frac{c6}{c7} \,. \tag{24}$$

This leads to

$$c7b = .019 \tag{25}$$

When one has the thin wall approximation removed, via b1 = 1.5, one does not see a difference in the ratios obtained.

For the 3$^{rd}$ potential system, which is intermediate between the 1$^{st}$ and 2$^{nd}$ potentials

if the b1 = b = 10 value is used, one obtains for when we have six dimensions

$$\alpha 1 := \frac{.373}{6} \tag{26}$$

As well as

$$V2(x) := \frac{1}{2} \cdot \frac{(\phi a(x))^2}{1 + .5 \cdot (\phi a(x))^3} \tag{27}$$

(When we have six dimensions)

$$C5b = 0.024 \tag{28}$$

(When we have three dimensions)

$$C7b = .016 \tag{29}$$

So, then one has C5b = .024, and C7b = .016 in the thin wall approximation

When b1 = 3 (non thin wall approximation)

$$C5b = .027 \tag{30}$$

(**Six dimensions**)

$$C7b = .02 \tag{31}$$

(**three dimensions**)

Summarizing, if

$$V1(x) := \frac{1}{2} \cdot (1 - \cos(\phi(x))) - \frac{1}{200} \cdot (\phi(x) - \pi)^2 \qquad = V1 \qquad (32)$$

$$V2(x) := \frac{1}{2} \cdot \frac{(\phi a(x))^2}{1 + .00001 \cdot (\phi a(x))^3} \qquad = V3 \qquad (33)$$

$$V2(x) := \frac{1}{2} \cdot \frac{(\phi a(x))^2}{1 + .5 \cdot (\phi a(x))^3} \qquad = V2 \qquad (34)$$

One finally obtains the following results, as summarized below

|  | b=b1 = 10 | b1 = 3 | b1 = 1 |
|---|---|---|---|
| V1 ( 6 dim) | C3b = .999 | No data | No data |
| V3 ( 6 dim) | C5b = 2.926E-3 | No data | C5b = same value |
| V3 ( 3 dim) | C7b = .019 | No data | C7b = same value |
| V2( 6 dim) | C5b = .027 | C5b = .024 | No data |
| V2 ( 3 dim) | C7b = .02 | C7b = .016 | No data |

Note that the table outlined above offers even more striking results. Namely that if one uses a higher 6 dimensional 'volume' element for initial nucleated space, that the agreement of Eq. (7) for a spatial six dimensional analysis as a starting point for the first potential will lead to an almost exact equality. Furthermore, if we use a normalization procedure as outlined in that appendix, and compare the ratios of both sides, that the relative slope of the scalar field will not be terribly important, in determining the relative contributions to both sides of Eq. (7) for $2^{nd}$ and $3^{rd}$ potentials.

[15] A. Beckwith arXIV math-ph/0411031: "An open question: Are topological arguments helpful in setting initial conditions for transport problems in condensed matter physics